\documentstyle[preprint,aps]{revtex}
\widetext
\draft
\tighten

\begin{document}

\title{On the difference between the charge-free and the charge-neutral
solutions of Maxwell equations}

\author{{\bf Andrew E. Chubykalo}\thanks{Corresponding author. E-mail:
andrew@cantera.reduaz.mx}}
\address{Escuela de F\'{\i}sica, Universidad Aut\'onoma de Zacatecas \\
Apartado Postal C-580\, Zacatecas 98068, ZAC., M\'exico}
\author{{\bf Hector A. M\'unera}}
\address{Centro Internacional de F\'{\i}sica, Apartado Aereo 251955,
Bogota D.C., Colombia}
\author{{\bf Roman Smirnov-Rueda}}
\address{Instituto de Ciencia de Materiales, CSIC, Cantoblanco, Madrid,
Spain}

\date{\today}

\maketitle

\baselineskip 7mm

\bigskip
\bigskip
\bigskip
\bigskip
\bigskip


\clearpage
\begin{abstract}
It is conventionally believed that solutions of so called ``free" Maxwell
equations for $\varrho=0$ (density of charge) describe the  free
electromagnetic field in empty space (if one considers the free field as a
field, whose flux lines neither {\it begin} nor {\it end} in a charge).  We
consider  three types of regions:  $(i)$ ``isolated charge-free" region
(where all electric fields, generated by charges outside that particular
region, {\it are zero}), for example, inside a hollow conductor of any
shape or in a free-charge Universe; $(ii)$ ``non-isolated charge-free"
region (where all electric fields, generated by charges outside that
particular region, {\it are not zero}) and $(iii)$ ``charge-neutral"
region (where  point charges exist but their algebraic sum is zero).   The
paper notes that there are two families of solutions: (1) In ``isolated
charge-free" regions electric {\it free field} does not exist {\it in the
context} of Maxwell's equations, but there may exist a {\it
time-independent} background magnetic field.  (2) In both
``charge-neutral"  and ``non-isolated charge-free" regions where the
homogeneous condition $\varrho=0$ also holds, Maxwell's equation for
electric field have non-zero solution, as in the conventional view, but
this solution is not {\it free field}.  We mention some implications
related to free-electromagnetic fields and the simplest charge-neutral
universe.

\end{abstract}

\section{Introduction}

It is well-known that the set of four Maxwell's equations (ME) [1,2]
describes different phenomena according to particular initial and boundary
conditions (BC). The authors of this note have independently found that
the set of solutions of ME may be larger than conventionally believed
[3-5]. As part of the process to establish BC for our generic problem, we
explore here the meaning of the solutions of ME in regions of space with
null charge density $(\varrho=0)$.

Conventionally, $\varrho=0$ represents ``empty space" (see, e.g., Purcell
[1], page 331). Under this condition, both Eqs. (3) and (4) (see below)
describe solenoidal fields, which imply that the electric  and magnetic
fields ({\bf E} and {\bf H}) in that region of space are transverse to the
{\it instantaneous} [6] direction of propagation. Moreover, since there
are no charges in such region, the electromagnetic wave corresponds to a
free field, whose flux lines neither begin nor end in a charge.

We want to argue here that such long-standing interpretation is not
completely consistent with the physics behind ME. The remainder of this
note is organized as follows: in Section II we critically revisit the
conventional interpretation to find that $\varrho=0$ leads to two families
of solutions. Section III explores some implications of our findings and
Section IV closes the paper.

\section{The conventional interpretation critically revisited}

In CGS units, Maxwell's equations are\footnote{Recently it was shown [4]
that one has to use total time-derivatives in (1)-(5) but one can do not
attach importance to this here. Recall that ${\bf E}={\bf D}$ and ${\bf
H}={\bf B}$ in vacuum in CGS units.}

\begin{eqnarray}
&& \nabla\cdot{\bf E}=4\pi\varrho,\\
&& \nabla\cdot{\bf H}=0,\\
&& \nabla\times{\bf H}=\frac{4\pi}{c}{\bf j}+\frac{1}{c}\frac{\partial
{\bf E}}{\partial t},\\
&& \nabla\times{\bf E}=-\frac{1}{c}\frac{\partial {\bf H}}{\partial t}.
\end{eqnarray}
Charge conservation is assured by the standard continuity condition:
\begin{equation}
\nabla\cdot{\bf j}+\frac{\partial\varrho}{\partial t}=0.
\end{equation}

We
consider  three types of regions:  $(i)$ ``isolated charge-free" region
(where all electric fields, generated by charges outside that particular
region, {\it are zero}), for example, inside a hollow conductor of any
shape or in a free-charge Universe; $(ii)$ ``non-isolated charge-free"
region (where all electric fields, generated by charges outside that
particular region, {\it are not zero}) and $(iii)$ ``charge-neutral"
region (where  point charges exist but their algebraic sum is zero).  Usually, one set\footnote{in (3) $\varrho{\bf V}={\bf j}_{cond}$}
 $\varrho = 0$ in (1) and (3) at the {\it whole} space (or in  ``isolated
charge-free" region, see $(i)$) and obtains {\it free} equations for {\it
free} field.  We argue here that this operation does not lead to
free-field solution of ME. For our argument, it is important to recall a
process  of obtaining Eqs.  (1) and (3).

We know that Gauss' law claims [1]: {\it The flux of the electric field
{\bf E} through any closed surface, that is, the integral $\oint {\bf
E}\cdot d{\bf a}$ over the surface, equals $4\pi$ times the total charge
enclosed by the surface}:
\begin{equation} \oint\limits_S {\bf E}\cdot d{\bf
a}=4\pi{\cal Q}=
4\pi\sum_i q_i=4\pi\int\limits_{\cal V}\varrho\, d\upsilon.
\end{equation}
We call this statement a {\it law} because it is {\it equivalent}   to
Coulomb's law and it could  serve equally well as the basic law of
electrostatic interactions, {\it after} charge and field have been {\it
defined}. Gauss' and Coulomb's laws are not two independent physical laws,
but the same law expressed in different ways. Looking back over a proof of
Eq.(6) in any textbook, we can see that it {\it hinged} on the {\it
inverse-square nature} of the interaction.  Thus this theorem (law) is
applicable {\it solely} to inverse-square field in physics. We stress
several aspects:

a) Coulomb's law is defined in terms of the individual $q_i$, so that the
expression for charge ${\cal Q}$ (Eq.(6)) in terms of charge density
$\varrho$ is only strictly valid as  a limit when a very large number of
charges is present. (Before we are criticized, we hasten to add that, of
course, $\varrho$ may be treated as $\delta$-function).

b) Gauss' law {\it only} applies to inverse-square fields, but they do not
need to be isotropic. Hence, it contains Coulomb's law, but it is somewhat
more general ([1]. p.24).

c) The right-hand-sides of Eq.(6) may be zero
in two different ways: (*) {\bf Charge-free condition}, ${\cal Q}=0$
when $q_i=0$, all ``$i$". (**) {\bf Charge-neutral condition},
${\cal Q}=0$ when $q_i\neq 0$, all ``$i$" independently.

Evidently, there is no reason to expect that {\bf E} on the {\it lhs} of
Gauss' law (6) should be the same for cases (*) and (**)
above.  Indeed, for an {\it isolated} charge-free region the {\it only}
solution is
\begin{equation}
{\bf E}=0,
\end{equation}
which simply means that a non-existing charge cannot produce an electric
field. Note that previous assertion is {\it qualitatively} different to
saying that there exist an electric field in the region that becomes zero
when ${\cal Q}=0$.

Let us remember now the Ostrogradsky-Gauss' theorem.  If this theorem
holds for any vector field, it certainly holds for {\bf E}:
\begin{equation}
\oint\limits_S {\bf E}\cdot d{\bf a}=\int\limits_{\cal
V}\nabla\cdot{\bf E}\, d\upsilon.
\end{equation}
Both Eq.(6) and Eq.(8) hold for any volume we care to choose - of any
shape, size, or location. Comparing them, we see that this can only be
true if at {\it every} point,
\begin{equation}
\nabla\cdot{\bf E}=4\pi\varrho.
\end{equation}
But we {\it always} must take into account that (because of origin of
(6)!) if $\varrho$ is zero in the  {\it isolated} charge-free region,
$\nabla\cdot{\bf E}$ must be zero because of {\bf E} is zero in the same
{\it isolated} charge-free region.

Now we can recall the origin of Eq.(3). Really, Maxwell found his famous
paradox (because of equation of continuity (5)):
\begin{equation}
\nabla\times{\bf H}=\frac{4\pi}{c}{\bf j}_{cond}+(?)
\end{equation}
and discovered what (?) must be:
\begin{equation}
\nabla\times{\bf H}=\frac{4\pi}{c}{\bf j}_{tot}=\frac{4\pi}{c}({\bf
j}_{cond}+{\bf j}_{disp}),
\end{equation}
\begin{equation}
\nabla\cdot{\bf j}_{tot}=\nabla\cdot{\bf j}_{cond}+\nabla\cdot{\bf
j}_{disp}=0,
\end{equation}
\begin{equation}
\nabla\cdot{\bf
j}_{disp}=-\nabla\cdot{\bf j}_{cond}=\frac{\partial\varrho}{\partial t}.
\end{equation}
Using (9) one obtains:
\begin{equation}
\nabla\cdot{\bf
j}_{disp}=\frac{1}{4\pi}\frac{\partial}{\partial t}\nabla\cdot{\bf
E}=\nabla\cdot\left(\frac{1}{4\pi}\frac{\partial{\bf E}}{\partial
t}\right).
\end{equation} General solution of this equation is
\begin{equation}
{\bf j}_{disp}=\frac{1}{4\pi}\frac{\partial{\bf
E}}{\partial t}+\nabla\times\left\{{\bf F}_1(x,y,z,t)\right\}+{\bf
F}_2(t)+{\bf const}.
\end{equation}

Maxwell (and others, following him) set the terms
\begin{equation}
\nabla\times\left\{{\bf
F}_1(x,y,z,t)\right\}+{\bf F}_2(t)+{\bf const}=0
\end{equation}
and as a results obtains  Eq.(3).
But {\it after} ({\bf
attention!}) obtaining Eq.(3) Maxwell {\it
at al} set following:

{\it In empty space, the terms with $\varrho$ and ${\bf
j}_{cond}=\varrho{\bf V}$ are zero,  and Maxwell's equations become}
\begin{eqnarray}
&& \nabla\cdot{\bf E}=0\\
&& \nabla\cdot{\bf H}=0\\
&& \nabla\times{\bf H}=+\frac{1}{c}
\frac{\partial{\bf E}}{\partial t}\\
&& \nabla\times{\bf E}=-\frac{1}{c}\frac{\partial{\bf H}}{\partial t}.
\end{eqnarray}

But if we set $\varrho=0$ at {\it whole} space  we must consider now (see
point (c) above) the {\bf charge-free condition} when $q_i=0$ for all $i$.
In other words, we obtain {\it isolated} charge-free region.   And as a
result we must obtain for this region
\begin{eqnarray}
&& \nabla\cdot{\bf
E}=0\\
&& \nabla\cdot{\bf H}=0\\
&& \nabla\times{\bf H}=(?)\\
&& \nabla\times{\bf E}=0
\end{eqnarray}
because {\bf E} has to  be zero in every point of this
region\footnote{the sense of (?) in Eq.(23) we explain in
Subsection A of Section III}. Of course,  in the {\it non-isolated}
charge-free region and in the charge-neutral region Eqs. (17)-(20) keep
their form but in this case we also cannot obtain an electric {\it free}
field, whose flux lines {\it neither begin nor end in a charge} (recall
the origin of Eq.(9)!).

We turn now to some implications of our interpretation.

\section{Implications of our interpretation}

\subsection{Isolated charge-free region}

Consider an isolated region ${\cal R}_0$ where no charges are present,
i.e. ${\cal Q}=0$, $\varrho=0$ everywhere. Eq. (7) applies, so that ${\bf
E} = 0$ everywhere in the whole space spanning ${\cal R}_0$. Assuming that
Maxwell's equations are valid in ${\cal R}_0$ it follows that magnetic
field {\bf H} may still exist, because Maxwell's Eq. (2) is completely
independent of $\varrho$. Indeed, in addition to the trivial solution
${\bf H}=0$, many other solutions of $\nabla\cdot{\bf H}=0$ are possible.
For instance, ${\bf H}=H_x{\bf i}+H_y{\bf j}+H_z{\bf k}$ with
$H_x=F_x(ct;y,z)$, $H_y=F_y(ct;x,z)$, $H_z=F_z(ct;x,y)$.

In a charge-free region Faraday's equation (4) reduces to
$\frac{\partial{\bf H}}{\partial t}=0$, hence {\bf H} is time-independent.
Our generic solution thus becomes $B_x=F(y,z)$, $B_y=F(x,z)$ and
$B_z=F(x,y)$, where we have noted that in isotropic region there is no
reason for the functional dependence to be different along arbitrary
orientations.

Finally, Amp\`ere's law (3) leads (see Eq.(15)) to
\begin{equation}
\nabla\times{\bf H}=\frac{4\pi}{c}{\bf j}_{\tt mag}
\end{equation}
where ${\bf j}_{\tt mag}$ may be some {\it magnetic} displacement current
density. Eq. (25) does not impose further constraints onto {\bf H}, but
rather defines the magnetic current ${\bf j}_{\tt mag}$. It may be
immediately verified that the continuity condition $\nabla\cdot{\bf
j}_{\tt mag}=0$ is fulfilled by all ${\bf j}_{\tt mag}$ defined by Eq.
(25). As an explicit example, let $B_x=F(y,z)=\sin[k(y+z)]$, et cyclicum.
Then, $j_x=(ck/4\pi)\{\cos[k(x+y)-\cos[k(x+z)]\}$ et cyclicum, where $k$
is in inverse length units.

Summarizing, in a charge-free region described by ME no electric field is
internally generated, but there {\it may} exist a time-independent magnetic
background.

\subsection{Non-isolated charge-free region}

Consider now a region ${\cal R}_0$ where no charges are present, ${\cal
Q}=0$, surrounded by a universe ${\cal U}$ where charges do exist. From
the superposition principle, total electric field in the region is ${\bf
E}({\cal R})={\bf E}({\cal R}_0)+{\bf E}({\cal U})={\bf E}({\cal U})$,
where ${\bf E}({\cal R}_0)$ denotes the field {\it internally} generated,
and ${\bf E}({\cal U})$ represents the field externally produced; from Eq.
(7), ${\bf E}({\cal R}_0)=0$. Likewise, for the total magnetic field in
the region, ${\bf H}({\cal R})={\bf H}({\cal R}_0)+{\bf H}({\cal U})$,
where ${\bf H}({\cal R}_0)$ is time-independent (see the discussion in
previous Subsection A).

It is thus clear that the electric field ${\bf E}({\cal R})$ existing
inside a charge-free region {\it is not a free field}; rather, it is
generated by charges outside the region. Of course, there is no
contradiction with Gauss' law (6) which refers to ${\bf E}({\cal U})$
entering and leaving the charge-free region.

\subsection{The simplest charge-neutral universe}

Consider a universe containing two equal charges of opposite sign. We can
easily obtain from ME with $\varrho=0$ different solutions \{${\bf
E}({\cal U}),{\bf H}({\cal U})$\}, depending upon the initial velocities
and separation of the charges.

Consider now a phenomenon that was unknown to Maxwell: charge
annihilation. What happens to the electric field ${\bf E}({\cal U})$ if
the charges meet to annihilate and form two photons? The obvious answer is
nothing, the electromagnetic field \{${\bf E}({\cal U}),{\bf H}({\cal
U})$\} continues its existence associated to the photons. None the less,
there is a difficulty because we are now in situation of ${\cal
Q}=0$\footnote{see point (c) in Section II}.

So, in a universe populated by two photons there are several fundamental
questions to answer. Firstly, do ME apply to them? Let us assume a
positive answer. Then, secondly, are we in a charge-free or in a
charge-neutral situation? Each possibility has different implications for
the inner structure of photons. If photons do not contain charge at all,
we are in a charge-free situation where the electric field has
disappeared: ${\bf E}(photons)=0$ (recall Eq. (7)). Hence, all information
about the photons must be contained in the time-dependent magnetic field
${\bf H}({\cal U})$. However, as discussed in Subsection A above, in a
charge-free region ${\bf H}({\cal R}_0)$ is time-independent, which means
that the field ${\bf H}({\cal U})$ is frozen in time at the moment of
annihilation.

Alternatively, if we are in a charge-neutral situation, then the
electromagnetic field \{${\bf E}({\cal U}),{\bf H}({\cal
U})$\} may continue to exist associated now to the two photons. But then,
it means that inside each charge-neutral photon there {\it must exist} at
least a hidden dipole! This interpretation nicely blends with the current
view from field theory that attaches electric dipole fields to photons.

\section{Concluding remarks}

In this paper we argued that a rigorous application of Gauss' law to the
solution of Maxwell's equations leads to the identification of two families
of solutions: {\it charge-free} and {\it charge-neutral}. This immediately
implies that electric {\it free field} does not exist {\it in the context}
of Maxwell's equations.

In an isolated charge-free vacuum, electric field does not exist, but
there {\it may} exist a time-independent background magnetic field. A
consideration of the simplest charge-neutral universe leads to some
interesting conjectures regarding the inner structure of photons.

\acknowledgments

The authors are indebted to Profs. Valeri Dvoeglazov (one of us (H.M.)
thanks his kind invitation to visit the School of Physics of the Zacatecas
University in the context of the CONACyT project No. 0270P-E), Stoyan
Vlaev, German Kalberman and Octavio Guzman  for helpful discussion and
critical comments.

\end{document}